This manuscript is a preprint of an article accepted for publication in *Social Science & Medicine*. The published version is available at https://doi.org/10.1016/j.socscimed.2026.119855

# AI Representations of Mental Health Stigma: Quantifying the Gap Between AI Estimations and Human Perceptions Across Gender Groups Using GPT-Based LLMs

Soojong Kim[1], Hye Min Kim[2], Kwanho Kim[3], and Dongdong Zhu[1]

[1] Department of Communication, University of California Davis, United States

[2] Department of Communication, University of Massachusetts Boston, United States

[3] Department of Media, College of Politics and Economics, Kyung Hee University, South Korea

## Author Note

Soojong Kim https://orcid.org/0000-0002-1334-5310

Correspondence concerning this article should be addressed to Soojong Kim, 1 Shields Ave, Kerr Hall #361, Davis, CA 95616, United States. Email: sjokim@ucdavis.edu

**Abstract**

The increasing use of artificial intelligence (AI) in mental health applications raises concerns about its potential to internalize and perpetuate societal biases. This research evaluates both disparities and misalignment in how large language models (LLMs), GPT-4 and GPT-5, represent perceptions of mental health across gender groups. The research comprises two studies. Study 1 applied widely adopted measures of Social Distance, Stereotypes, and Attitudes Toward Seeking Help to assess whether the AI models depict systematic gender differences. Study 2 compared AI estimations of Personal, Self, and Perceived Stigma with large-scale real-world human data from the Healthy Minds Study ($N$ = 76,406) to evaluate alignment between AI estimations and human perceptions. Results from Study 1 showed the AI models consistently portrayed men as having the greatest desire for social distance from someone with mental health problems, relatively higher Stereotypes scores, and the least favorable attitudes toward help-seeking, compared with women and nonbinary individuals. Study 2 confirmed that this ordinal pattern partially aligned with human-reported data for Personal and Self Stigma. However, the AI models systematically underestimated Personal Stigma and Perceived Stigma while substantially overestimating Self Stigma for all gender groups. These findings reveal that, although the AI models have learned relative gender-based patterns reflecting human experiences to a certain extent, they function as distorted portrayals, highlighting the need for rigorous, context-specific validation against real-world data to ensure AI's safe and equitable deployment.

**AI Representations of Mental Health Stigma: Quantifying the Gap Between AI Estimations and Human Perceptions Across Gender Groups Using GPT-Based LLMs**

The rise of artificial intelligence (AI) is transforming public health and healthcare, bringing both challenges and opportunities to improve health surveillance, disease diagnosis, healthcare access, health communication, and data-driven decision-making (Matheny et al., 2020; Panteli et al., 2025). A growing number of AI applications now focus on mental health, ranging from conversational agents that provide social support to AI models that analyze human behaviors and linguistic markers of distress (Lee et al., 2021; Rajpurkar et al., 2022). AI-powered platforms are especially appealing to young people, a demographic that represents early adopters of new technologies and experiences an increasing prevalence of mental health challenges (Keyes et al., 2019; Vigo et al., 2016). The anonymity, accessibility, and affordability of AI-powered tools provide a low-barrier entry point for individuals who might otherwise hesitate to seek professional help (Luxton, 2020; Rodríguez-Rivas et al., 2022).

However, concerns remain regarding the accuracy, fairness, and disparities of AI in representing human experiences and perceptions of mental health (Carr, 2020; Timmons et al., 2023). Biases and misrepresentations in AI models in mental health that power interactive systems, conversational agents, and technologies designed to analyze human behavior and linguistic data could lead to multiple levels of harm. First, they can lead to biased and inaccurate diagnosis and advice, potentially less encouraging or even discouraging people with mental health concerns from seeking necessary medical and social support. Second, they can shape users' perceptions of how mental illness is viewed by others, possibly causing them to overestimate public stigma and stereotypes, thereby creating an additional barrier to help-seeking. Lastly, at a broader level, such biases can perpetuate structural inequities by amplifying underrepresentation or misrepresentation of specific populations within digital health systems (Gianfrancesco et al., 2018; Seyyed-Kalantari et al., 2021; Timmons et al., 2023).

The concepts of stigma (e.g., negative stereotypes toward mental health problems, unwillingness to engage with people experiencing mental health issues) and attitudes toward help-seeking are central to understanding barriers to mental healthcare. Research has shown that stigma is a major factor discouraging individuals from accessing professional help, often due to fears of shame and negative social judgment (Clement et al., 2015).

It should be noted that stigma is not a unitary construct. The literature on mental illness stigma distinguishes public stigma from self-stigma and differentiates cognitive, affective, and behavioral components of stigmatizing responses (Corrigan et al., 2003; Fox et al., 2018). Public stigma refers to the ways in which members of the general public stigmatize people with mental illness, whereas self-stigma refers to the internalization of devaluation by people who may themselves experience mental health difficulties (Corrigan, 2000; Corrigan & Watson, 2002). Within public stigma, Corrigan et al. (2003) further elaborate that stereotypes reflect negative beliefs, prejudice reflects negative emotional reactions, and discrimination reflects behavioral responses such as rejection, avoidance, or social distance.

As AI systems increasingly serve as central interaction points for mental health support, how AI estimates or “portrays” these constructs in society becomes more important. For instance, if AI models depict that people have overly high levels of stigma and interact with their users based on these estimations, it could worsen users’ concerns about how their mental illness would be perceived by others, amplifying the very barriers that public health initiatives seek to overcome (Corrigan et al., 2012).

This challenge is further complicated by gender disparities. Prior research suggests that gender differences in stigma and help-seeking may arise from social norms governing emotional expression, vulnerability, and autonomy (Addis & Mahalik, 2003; Gonzalez et al., 2005; Holzinger et al., 2012; Mackenzie et al., 2006). Masculinity norms often reward self-reliance and emotional control, which may increase resistance to formal help-seeking and intensify stigma (Addis & Mahalik, 2003; Möller-Leimkühler, 2002; Seidler et al., 2016). By contrast, women often face fewer social penalties for emotional disclosure and support-seeking, which may contribute to more favorable attitudes toward professional help and higher rates of service use (Clement et al., 2015; Gonzalez et al., 2005; Mackenzie et al., 2006). These tendencies are socially patterned rather than biologically fixed (Courtenay, 2000; Möller-Leimkühler, 2002).

The present research therefore focuses on gender because gender-linked differences in public stigma, self-stigma, and help-seeking are well established in the literature and are directly relevant to the use of AI in mental health contexts. We treat this focus as a targeted first step rather than a comprehensive account of all socially structured variation in AI portrayals of mental health stigma.

If an AI model recognizes and reproduces these differences, its function becomes dependent on the gender of users or people being analyzed by AI. This creates a dynamic whose direction is not immediately clear, since the model may either help close gender gaps or contribute to widening them. For example, would it encourage men to seek professional help regarding a mental health problem more than women (to offset men's negative attitudes) or less (to accommodate their preferences and maintain user engagement)? AI models may also attend selectively to different aspects of an individual's mental health experience depending on gender (e.g., Plaza Del Arco et al., 2024).

These concerns motivate the need to examine how AI models portray mental health stigma and help-seeking across gender groups. Additional background on the promise, risks, and validation challenges of AI in mental health is provided in the Supplementary Online Material (SOM) Section 1. The next section reviews algorithmic bias in mental health applications and prior work on gendered representations in AI outputs.

## Algorithmic Bias in Mental Health AI

Biases in AI models are a substantial barrier to the safe and equitable deployment of AI in mental health (Agarwal et al., 2023; Carr, 2020; Graham et al., 2019; Lee et al., 2021; Straw, 2020; Timmons et al., 2023). AI models learn patterns from the data on which they are trained, and if the data reflects existing societal stereotypes and inequities, the models can easily learn and encode them (e.g., Koenecke et al., 2020; Obermeyer et al., 2019; Williams, Brooks, et al., 2018). However, biases do not originate solely from training data. Biases can infiltrate the AI lifecycle at every stage, from unrepresentative data collection and biased labels to flawed model construction and evaluation (Agarwal et al., 2023; I. Chen et al., 2021; Luxton, 2020; Timmons et al., 2023). The result is AI systems that may perform inequitably across social groups, worsening care disparities for already underserved populations.

Gender bias has become a pressing concern in the development and application of AI for mental health. Substantial evidence has documented gender-based biases and discrimination in algorithms and AI in other domains, such as facial recognition technologies (Menezes et al., 2021), web search systems (Makhortykh et al., 2021; Vlasceanu & Amodio, 2022), housing (Asplund et al., 2020), and employment (L. Chen et al., 2018; Imana et al., 2021). More broadly, recent research shows that even LLMs presented as explicitly unbiased can still form biased associations (Bai et al., 2025). In mental-health-related contexts, Lauderdale et al. (2025) found that LLMs evaluating vignette characters with borderline personality disorder (BPD) exhibited gender bias, showing more negative reactions and greater public stigma directed toward a woman with BPD. Other studies similarly suggest that LLM outputs can vary systematically by gender and mental health condition. For example, Schnepper et al. (2025) found that an LLM assumed a greater impairment in mental health-related quality of life for men compared with women when evaluating identical case vignettes for anorexia and bulimia. Lin et al. (2022) demonstrated that language models are consistently less likely to associate men with mental health conditions compared with women, and this disparity widens in sentences describing treatment-seeking behavior. Chansiri et al. (2024) reported that LLMs over-associate men with narcissistic personality disorder and women with BPD. Also, Noguero et al. (2023) analyzed AI algorithms trained to detect anorexia nervosa on social media and found that the models demonstrated higher false negative rates for female users than male users. The implications of these biased patterns from the perspective of AI explainability are discussed further in the SOM Section 1.

The present research investigates how large language models (LLMs), one of the most rapidly evolving and widely adopted technologies in AI, represent perceptions of mental health across different gender groups. It aims to identify the specific gender-based patterns present in AI estimations and to determine how much these patterns are aligned with or deviated from real-world human perceptions.

Here, it is important to distinguish between two forms of potential bias of AI. The first is “disparity,” wherein AI models produce systematically different outcomes or associations depending on social groups. A model may generate different outputs for specific groups, potentially due to differences in the internalized assumptions and learned associations (Garg et al., 2018; Obermeyer et al., 2019). Disparities in AI outcomes may or may not reflect underlying differences in real-world outcomes, and the absence of such disparities does not necessarily indicate accuracy or fairness (Hardt et al., 2016). The second, and more intricate, form is "misalignment," or an imperfect mirroring of reality. A model may be misaligned if its estimations fail to capture real-world human experiences and perceptions accurately, not only through over- or underestimation but also by misrepresenting the true direction or magnitude of real-world group differences (Amodei et al., 2016; Gabriel, 2020). These two forms of bias do not necessarily co-occur. A model may generate outcomes showing disparity between social groups that nonetheless reflect real-world observations (disparity with alignment). On the other hand, a model may produce outcomes showing no meaningful group differences, even when such equivalence does not reflect the reality (parity with misalignment). Although it is known that certain statistical criteria requiring group parity cannot be satisfied at the same time as accuracy-based criteria leading to perfect alignment (Chouldechova, 2017; Kleinberg et al., 2017), it is still meaningful to evaluate the levels and patterns of these imperfections in important real-world AI applications such as mental health.

The present research is structured to evaluate both forms of potential representational AI bias. The research comprises two sequential studies designed to analyze these representations in depth. In Study 1, LLMs were prompted to generate a large volume of responses reflecting stigma toward mental illness and attitudes toward help-seeking across different gender groups. This study was designed to identify the existence and patterns of disparity in the AI portrayals of mental health attitudes across gender groups. Study 2 then moves beyond identification to assess alignment, by quantifying the specific nature and magnitude of these AI estimations against large-scale real-world human data. In this study, AI-estimated stigma levels were directly compared with population-level survey data from the Healthy Minds Study (HMS), one of the largest annual surveys of mental health in the U.S. postsecondary population (Healthy Minds Network, 2023). The focus on gender allowed us to compare AI estimations with sufficiently large corresponding groups in the HMS data.

The prior studies show that AI models encode gender-based disparities and produce outcomes that reflect broader societal biases. Extending this pattern to the domain of stigma, and considering the tendency of AI models to encode traditional masculinity norms that associate men with emotional suppression and resistance to help-seeking (Addis & Mahalik, 2003; Lin et al., 2022; Plaza-del-Arco et al., 2024; Seidler et al., 2016), Study 1 was guided by the following research question and hypothesis:

**RQ1** (Disparity in Study 1). Do AI models exhibit gender-group disparities in their estimations of Social Distance, Stereotypes toward people with mental health problems, and Attitudes Toward Seeking Professional Help?

**H1** (Disparity in Study 1). AI models will exhibit gender disparities such that men are estimated to show higher Social Distance, stronger Stereotypes toward people with mental health problems, and less favorable Attitudes Toward Seeking Professional Help than women and nonbinary individuals.

Little research has examined how AI models portray the stigma and attitudes of gender groups or attempted to quantify misalignments. The present research addresses this gap by comparing AI-estimated attitudes of gender groups toward mental health and by quantitatively evaluating LLMs' estimations of mental health stigma against real-world survey data. Because Study 2 used the HMS data as an empirical benchmark, we did not assume in advance that a specific gender-group ordering would necessarily appear in the human-reported data for each stigma measure. Therefore, Study 2 was guided by non-directional research questions that first examine whether AI estimations correspond to the gender-group ordering observed in the HMS data and then quantify the magnitude and direction of AI-human discrepancies:

**RQ2** (Misalignment in Study 2: Ordering alignment). To what extent do AI estimations of Personal Stigma, Self Stigma, and Perceived Stigma show the same ordering of gender groups as the HMS data?

**RQ3** (Misalignment in Study 2: Magnitude and direction). For each gender group and each stigma measure, what is the magnitude and direction of the discrepancy between AI estimations and human perceptions?

# Methods

## AI-Estimated Mental Health Attitudes Across Gender Groups (Study 1)

The primary objective of Study 1 was to investigate whether leading LLMs exhibit systematic disparities in their depiction of mental health perceptions and attitudes across different gender groups.

### *Measures*

The constructs examined in Study 1 capture related but distinct aspects of public responses to mental illness: Social Distance as a behavioral expression of rejection or discrimination, Stereotypes as the cognitive component of public stigma, and Attitudes Toward Seeking Help as a related treatment-seeking outcome closely tied to stigma processes (Corrigan et al., 2003; Fox et al., 2018).

**Social Distance**. We used six items adapted from the social distance measures commonly used in stigma research (Hoffner & Cohen, 2018; Martin et al., 2000; Pescosolido et al., 2021; S.-L. L. Williams et al., 2018). The AI models were prompted to estimate how willing members of a given gender group in the U.S. (male, female, or nonbinary individuals aged 18 or older) would be to engage in a given behavior involving a person with a mental health problem, such as "make friends with the person." Items were rated on a four-point scale from "definitely willing" (1) to "definitely unwilling" (4). Responses of 1 or 2 were recoded as 0, and responses of 3 or 4 were recoded as 1. The six recoded responses were summed to compute an overall score ranging from 0 to 6. Higher scores indicated less willingness and thus greater perceived social distance from people with a mental health problem.

**Stereotypes**. We used seven items adapted from an established measure (Hoffner & Cohen, 2015, 2018) to assess negative stereotypical beliefs, such as "A person with depression is dangerous." The AI models rated each statement on a seven-point Likert scale from "strongly disagree" (1) to "strongly agree" (7). Three items were reverse-coded, and ratings were averaged to compute an overall score ranging from 1 to 7. Higher scores indicated stronger endorsement of negative stereotypes toward people with mental illness.

**Attitudes Toward Seeking Help**. We used ten items from the Attitudes Toward Seeking Professional Psychological Help Scale–Short Form (ATSPPH-SF; Elhai et al., 2008; Fischer & Farina, 1995), such as "If I believed I was having a mental breakdown, my first inclination would be to get professional attention." The AI models rated each item using a four-point scale from "disagree" (0) to "agree" (3). Five items were reverse-coded, and all ten ratings were summed to compute an overall score ranging from 0 to 30. Higher scores indicated more positive attitudes toward seeking professional help. Further measurement details are provided in the SOM Section 2.1, and example prompts are presented in Table S1.

### ***Generation of LLM Responses***

For each construct and gender group, GPT-4 (gpt-4o-2024-05-13) generated 1,000 independent, stateless responses with randomized item order and temperature set to the midpoint of the model's available range. GPT-5 mini (gpt-5-mini-2025-08-07) was also used to generate responses for the robustness analysis. API procedures, parameter settings, reasoning configuration, and model-selection rationale are provided in the SOM Section 2.2.

**Comparison Between AI Estimation and Human Perception (Study 2)**

Study 2 aimed to compare AI-estimated perceptions with actual human perceptions using large-scale real-world survey data. We used human responses from the HMS, one of the largest and most comprehensive population-level surveys of mental health among postsecondary students in the U.S. (Healthy Minds Network, 2023). We requested and received the most recent available HMS dataset from the Healthy Minds Network, which leads the HMS, in May 2024. Our analysis was based on the 2022-2023 academic year dataset, the most recent dataset available to us at the time, which included responses from 76,406 respondents. The mean age of respondents was 23.28 years (*SD* = 7.19). The present analysis focused on three gender categories—female (68.52%), male (25.74%), and nonbinary (2.59%)—to align the human-response data with the categories examined in the AI estimation task, capturing 97% of respondents. More details on gender identity assessment are reported in the SOM Section 2.3.

***Measures***

Study 2 used three three-item stigma measures from the HMS (Healthy Minds Network, 2023). Personal Stigma assessed an individual's own stigmatizing attitudes toward others with a mental illness, such as "I feel that receiving mental health treatment is a sign of personal failure." Self Stigma assessed the extent to which an individual would feel embarrassed or ashamed about having a mental illness themselves, such as "I would feel inadequate if I went to a therapist for psychological help." Perceived Stigma assessed an individual's perception of the stigmatizing attitudes held by others, such as "Most people feel that receiving mental health treatment is a sign of personal failure." Personal Stigma and Perceived Stigma used six-point scales from strongly agree (1) to strongly disagree (6), whereas Self Stigma used a five-point scale from strongly disagree (1) to strongly agree (5). Items were coded or reverse-coded as needed so that higher values consistently indicated greater stigma, and the three items for each measure were averaged to create overall scores. In the human HMS sample, the internal consistency was $\alpha = .73$ for Perceived Stigma, .93 for Self Stigma, and .74 for Personal Stigma. We retained the HMS labels of the constructs while recognizing that Personal Stigma aligns most closely with respondents' own public-stigma-related beliefs, Perceived Stigma with perceived stigma, and Self Stigma with treatment-related internalized or self-stigma (Corrigan et al., 2003; Fox et al., 2018).

### *AI Estimation Procedure*

Study 2 followed the Study 1 procedure for response generation. The prompts used the exact HMS item wording and response scales and asked each model to estimate the responses of a male, female, or nonbinary student enrolled at a U.S. college or university, living in the U.S., and aged 18 years or older. This student-specific reference population aligned the AI estimation task with the HMS sample. An example prompt is presented in Table S1, and further procedural and model details are provided in the SOM Section 2.5.

To analyze and compare the data, we first calculated the mean for each of the three stigma measures by gender group, using both the AI-generated data and the human-reported data from the HMS. These means were then directly compared to identify discrepancies between the AI estimations and human perceptions. To quantify the magnitude of these discrepancies, we also calculated AI bias for each stigma measure and gender group. The bias score was defined as the difference between the AI-estimation mean and the human-report mean (i.e., AI bias = $M_{AI}$ - $M_{Human}$). A positive bias score indicates that an AI model overestimates the level of stigma for that group compared with human reports, while a negative score indicates underestimation. Because the stigma measures differed in response range, we also calculated range-standardized AI bias: standardized AI bias = ($M_{AI}$ - $M_{Human}$) / (scale maximum - scale minimum). This approach preserves the direction of bias while making the magnitude of discrepancies more comparable across measures, and was used for visualization purposes.

## Methodological Considerations

### *Analysis of Group-Level Differences*

The Mann-Whitney test was used to analyze differences in constructs between gender groups estimated by an AI model, between gender groups in the human-reported data, and between AI estimations and human reports for the same gender group. This is a nonparametric test that does not rely on the assumption of normality. The present research reports the Mann–Whitney *U* scores, *p*-values, and Common-Language Effect Sizes (CLES; *f*) for this test (the Mann-Whitney *U* statistics are reported in the SOM due to their large numerical values). The CLES represents the probability that a randomly selected score from one group will be higher than a randomly selected score from another group, ranging from 0 to 1. CLES values greater than 0.5 indicate that the first group tends to have higher scores than the second, whereas values less than 0.5 indicate the opposite. When comparing different gender groups within AI-estimated results or within human-reported responses, Bonferroni adjustment was applied to reduce the risk of Type I error due to multiple comparisons, and the adjusted *p*-values ($p_{adj}$) are reported.

To compare AI-estimated scores across gender groups and with human-reported scores, we examined the mean scores for each construct. For completeness, we also report inferential statistics, but these should be interpreted with caution. Because this research involves simulated data using AI, the resulting *p*-values for AI estimations primarily reflect the number of model runs rather than the true variability of the underlying "population," that is, all possible LLM responses. In such simulations, even trivial differences can reach statistical significance if repeated a sufficient number of times. Therefore, the key insight lies not in whether group differences in AI estimations are statistically significant, but in the consistency and stability of an AI model's estimations across measures, items, and contexts. The reported statistical results thus serve a descriptive rather than inferential purpose, illustrating that the observed mean differences represent reproducible features of an AI model's internal representations rather than random noise.

***Approach for LLM Response Generation***

In both Study 1 and Study 2, we instructed the AI models to estimate a gender group's responses to specific survey questions. We used established survey measures as standardized measurement tools rather than assuming that conventional human psychometric properties automatically transfer to model outputs. This design enabled comparisons across gender groups and, in Study 2, direct comparison with the HMS human responses. The labels "female individuals," "male individuals," and "nonbinary individuals" were used as operational categories to maintain parallel wording across prompts. These labels were also intended to match the gender categories used in the HMS survey, which assessed gender identity using response options that included "male," "female," and "gender non-binary." We used zero-shot prompting without examples of desired answers, preserving correspondence with the original survey items and reducing the possibility that researcher-provided examples would shape the responses. This task evaluates model portrayals under controlled survey prompting rather than model behavior in naturalistic conversation.

We examined whether 1,000 iterations were sufficient using running-mean convergence diagnostics. We also repeated the GPT-4 analyses under lower and higher temperature settings, a paraphrased prompt template, and a role-framing prompt. Detailed rationale is provided in the SOM Section 2.6, and the procedures and results of the robustness analyses are reported in the SOM Section 3.

## Results

### AI-Estimated Mental Health Attitudes Across Gender Groups (Study 1)

To address RQ1 and test H1, we examined whether the AI models exhibited gender-group disparities across social distance, stereotypes, and attitudes toward seeking professional help, and whether the direction of these disparities was consistent with the hypothesized ordering.

The results showed consistent gender-based patterns in the AI estimations of mental health perceptions, which are summarized in Figure 1. Overall, the AI models depicted male individuals as perceiving greater social distance to people with mental illness, higher levels of stereotypes, and less positive attitudes toward seeking help, compared with female and nonbinary individuals. While the section below details the primary results from GPT-4 (Figures 1a-1c), these patterns were found to be robust in the additional analysis using GPT-5 (Figure 1d).

First, disparities among gender groups emerged for Social Distance (Figure 1a). GPT-4 estimated that male individuals ($M = 4.02$, $SD = 0.58$) perceive the greatest social distance, significantly more than female individuals ($M = 3.87$, $SD = 0.51$; $p_{adj} < .001$, $f = 0.56$) and nonbinary individuals ($M = 3.59$, $SD = 0.65$; $p_{adj} < .001$, $f = 0.66$). Female individuals were also estimated to perceive significantly greater social distance than nonbinary individuals ($p_{adj} < .001$, $f = 0.61$).

An item-level analysis showed this pattern (male > female > nonbinary) was consistent. For instance, on the item “make friends with the person” (“Friend” in Figure 1a), the estimated unwillingness was highest for male individuals ($M = 1.94$, $SD = 0.46$), followed by female individuals ($M = 1.82$, $SD = 0.46$), and lowest for nonbinary individuals ($M = 1.69$, $SD = 0.51$). This specific ordering of means was consistent for all six items, even including the item “have the person marry into their family” (“Marry” in Figure 1a), which exhibited a strong ceiling effect. For this item, GPT-4 estimated all gender groups would be “definitely unwilling” (a score of 4) in most cases (male: 99.1%, female: 98.4%, nonbinary: 96.0%), but the proportion of these responses still followed the same ordinal pattern.

Second, the AI estimations for Stereotypes (Figure 1b) followed the same trend. Male individuals were depicted as holding relatively stronger stereotypes ($M = 2.97$, $SD = 0.30$) than female individuals ($M = 2.75$, $SD = 0.22$; $p_{adj} < .001$, $f = 0.72$) and nonbinary individuals ($M = 2.60$, $SD = 0.28$; $p_{adj} < .001$, $f = 0.83$). GPT-4 also estimated that female individuals hold significantly stronger negative stereotypes than nonbinary individuals ($p_{adj} < .001$, $f = 0.67$). However, all three group means were below the scale midpoint of 4 on the seven-point scale, which indicates that the AI model leaned toward disagreement with stereotypical statements in absolute terms, despite the relative differences across gender groups.

This overall pattern for Stereotypes was consistent at the item level. For instance, for the item “A person with depression is dangerous” (“Dangerous” in Figure 1b), the mean agreement estimated by the AI model was highest for male individuals ($M = 2.09$, $SD = 0.34$), lower for female individuals ($M = 2.00$, $SD = 0.18$), and lowest for nonbinary individuals ($M = 1.68$, $SD =$ 0.47). This specific ordering (male > female > nonbinary) was observed for six of the seven Stereotypes items. The only item that deviated was "A person with depression is able to maintain a regular job" ("Regular job" in Figure 1b), where although male individuals' reverse-coded mean score ($M = 4.07$, $SD = 0.74$) was still higher than female individuals' ($M = 3.89$, $SD = 0.73$), nonbinary individuals' mean ($M = 4.08$, $SD = 0.75$) was estimated to be similar to male individuals'.

Lastly, the AI estimations of Attitudes Toward Seeking Help also differed across gender groups (Figure 1c). GPT-4 consistently estimated less positive attitudes toward professional help-seeking among male individuals, which is aligned with the stronger stigma reflected in the other two constructs (Social Distance and Stereotypes). The AI model estimated significantly less positive attitudes toward seeking professional help among male individuals ($M = 19.54$, $SD = 3.56$) compared with female individuals ($M = 24.53$, $SD = 1.61$; $p_{adj} < .001$, $f = 0.09$) and also compared with nonbinary individuals ($M = 25.27$, $SD = 2.02$; $p_{adj} < .001$, $f = 0.07$). The model also estimated that female individuals had significantly less positive attitudes toward seeking help compared with nonbinary individuals ($p_{adj} < .001$, $f = 0.37$).

This pattern was robust at the item level. The AI model estimated male individuals to have the lowest (least positive) mean score on all 10 items. The gap between male individuals and the other two groups was pronounced for all items. For example, on the item "Self solve first" (i.e., "A person should work out his or her own problems; getting psychological counseling would be a last resort."), the mean for male individuals ($M = 1.52$, $SD = 0.75$) was substantially lower than for female ($M = 2.27$, $SD = 0.47$) and nonbinary individuals ($M = 2.56$, $SD = 0.50$). In contrast, the pattern of attitudes between female and nonbinary individuals was more mixed at the item level. While nonbinary individuals had a significantly more positive overall score, this did not hold for every item.

To confirm the robustness of these findings, we conducted the same analysis using GPT-5. As illustrated in Figure 1d, the results were consistent with those from GPT-4, replicating the same gender-based ordinal patterns across all three constructs.

**Comparison of AI Estimations and Human Perceptions (Study 2)**

Study 2 compared the AI estimations of mental health stigma with the human perceptions of stigma in the real world (Figure 2). To address RQ2 and RQ3, we focused on the extent to which AI estimations aligned with the ordinal patterns observed in human perceptions and on the magnitude and direction of discrepancies between AI estimations and human perceptions.

Consistent with the results of Study 1, the AI models estimated higher levels of Personal Stigma (an individual's own stigmatizing attitudes toward others with a mental illness) and Self Stigma (the extent to which an individual would feel embarrassed or ashamed to have a mental illness themselves) among male individuals than female and nonbinary individuals. The results of Study 2 also reveal discrepancies in magnitude between AI estimations and human perceptions of stigma but show that the AI estimations may reflect the ordinal patterns observed in human perceptions. While this section details the primary results from GPT-4 (Figure 2), these patterns were largely replicated in the validation analysis using GPT-5 (Figure S5 in the SOM).

The comparison revealed two distinct types of findings, which are presented in detail below. First, we analyzed the discrepancies in magnitude, by comparing the absolute mean stigma scores from the AI estimations with the human-reported means for each gender group. Second, we evaluated the alignment of ordinal patterns, by comparing the relative differences between gender groups, e.g., whether the group with the highest stigma in the human data was also estimated to have the highest stigma by AI.

GPT-4 overestimated the level of Self Stigma for all three gender groups (Figures 2b, 2e, 2g). This overestimation was substantial across all groups: female ($M_{AI} = 2.80$, $M_{Human} = 1.89$; $p < .001$, $f = 0.79$), male ($M_{AI} = 2.91$, $M_{Human} = 2.19$; $p < .001$, $f = 0.73$), and nonbinary individuals ($M_{AI} = 2.44$, $M_{Human} = 1.68$; $p < .001$, $f = 0.78$).

Contrarily, the AI model underestimated the levels of Personal Stigma (Figures 2a, 2d, 2g) for all groups. This degree of underestimation was greatest for male individuals ($M_{AI} = 1.08$, $M_{Human} = 1.77$; $p < .001$, $f = 0.28$) but was also significant for female ($p < .001$, $f = 0.33$) and nonbinary individuals ($p < .001$, $f = 0.40$). However, it should be noted that AI estimates of Personal Stigma were concentrated near the lower end of the response scale. Therefore, between-group differences for this construct should be interpreted with caution as reflecting a compressed, floor-adjacent ordering rather than strong evidence of substantial group separation. The full item-level response distributions are presented in the SOM Section 4.

A significant underestimation was also found for Perceived Stigma (an individual's perception of the stigmatizing attitudes held by others): male ($M_{AI} = 2.50$, $M_{Human} = 2.99$; $p < .001$, $f = 0.38$) and female individuals ($M_{AI} = 2.49$, $M_{Human} = 2.92$; $p < .001$, $f = 0.41$). GPT-4's estimated mean of Perceived Stigma for nonbinary individuals ($M_{AI} = 2.78$) was also lower than the human mean ($M_{Human} = 2.98$; $p = .009$, $f = 0.47$).

Despite these differences in absolute magnitude, the AI model's estimations partially reflected the relative differences and ordinal patterns observed in human perceptions. For Personal Stigma, the HMS data showed that male individuals ($M = 1.77$, $SD = 1.03$) reported significantly higher Personal Stigma than female individuals ($M = 1.44$, $SD = 0.78$; $p_{adj} < .001$, $f$

= 0.60) and nonbinary individuals (*M* = 1.22, *SD* = 0.55; $p_{adj} < .001$, *f* = 0.67). The AI model's estimations captured this same ordinal pattern: male individuals (*M* = 1.08, *SD* = 0.21) were depicted by the AI model as holding significantly higher Personal Stigma than female individuals (*M* = 1.01, *SD* = 0.07; $p_{adj} < .001$, *f* = 0.56) and nonbinary individuals (*M* = 1.01, *SD* = 0.05; $p_{adj} < .001$, *f* = 0.56) (Figures 2a, 2d).

A similar alignment was found for Self Stigma. In the human-reported data, male individuals (*M* = 2.19, *SD* = 1.01) reported the highest levels of stigma, greater than female individuals (*M* = 1.89, *SD* = 0.90; $p_{adj} < .001$, *f* = 0.59) and nonbinary individuals (*M* = 1.68, *SD* = 0.87; $p_{adj} < .001$, *f* = 0.66). GPT-4 estimations, although they were uniformly higher than human perceptions, mirrored this exact order: male individuals (*M* = 2.91, *SD* = 0.72) were estimated to have the highest Self Stigma, significantly higher than female individuals (*M* = 2.80, *SD* = 0.73; $p_{adj} = .003$, *f* = 0.54) and nonbinary individuals (*M* = 2.44, *SD* = 0.68; $p_{adj} < .001$, *f* = 0.68) (Figures 2b, 2e).

However, this alignment was less clear for Perceived Stigma. In the human responses, male individuals (*M* = 2.99, *SD* = 1.22) reported significantly higher Perceived Stigma than female individuals (*M* = 2.92, *SD* = 1.23; $p_{adj} < .001$, *f* = 0.52), while nonbinary individuals (*M* = 2.98, *SD* = 1.24) were not significantly different from male individuals ($p_{adj} = 1.000$, *f* = 0.51). On the other hand, the AI model estimated that, compared with nonbinary individuals (*M* = 2.78, *SD* = 0.76), both female (*M* = 2.49, *SD* = 0.62; $p_{adj} < .001$, *f* = 0.40) and male individuals (*M* = 2.50, *SD* = 0.53; $p_{adj} < .001$, *f* = 0.39) had significantly lower Perceived Stigma, while estimating male and female individuals as nearly identical ($p_{adj} = 1.000$, *f* = 0.50) (Figures 2c, 2f).

Finally, we verified these patterns using GPT-5 (Figure S5 in the SOM). This additional analysis confirmed that the model exhibited similar systematic biases, substantially overestimating Self Stigma while underestimating Personal and Perceived Stigma. Regarding gender-based patterns, the model preserved the primary trends found in its predecessor: male individuals were consistently estimated to hold the highest levels of Personal and Self Stigma, whereas nonbinary individuals were estimated to perceive the highest level of stigma from others, despite a shift in the relative ranking between female and nonbinary individuals for Self Stigma.

Additional sensitivity checks using GPT-4 also supported the robustness of the main findings. Across these checks, the main Study 1 finding that male individuals were estimated to show greater Social Distance, stronger Stereotypes, and less positive Attitudes Toward Seeking Professional Help was consistently reproduced, and the main Study 2 findings regarding partial ordinal alignment and systematic discrepancies in magnitude between AI estimations and human responses were largely retained. The detailed results of these sensitivity checks are presented in the SOM Section 3.

As a descriptive index, we calculated Cronbach's alpha for the AI-generated responses, while recognizing that these coefficients should not be interpreted in the same way as psychometric reliability estimates from human responses. The coefficients varied across constructs and gender groups, with values ranging from near zero to .80 in Study 1 and from .32 to .79 in Study 2, indicating that the AI-generated composites did not exhibit uniformly strong scale-level coherence. All coefficients are reported in the SOM Section 4.

## Discussion

This research investigated how GPT-based LLMs portray perceptions of mental health problems and attitudes toward help-seeking across different gender groups. It also compared the AI estimations against actual human perceptions. Related work has used different methodological approaches to examine biased associations and stigma in LLM outputs more directly (e.g., Bai et al., 2025; Lauderdale et al., 2025), whereas the present research focused on disparities in prompted estimations across gender groups and their misalignment with human-reported perceptions. The results show that, under the present role-based prompting strategy, the AI models produced consistent, gender-differentiated representations of mental health attitudes that partially aligned with human-reported perceptions, while also showing substantial discrepancies in magnitude.

Across two studies, the GPT-based AI models consistently depicted men as holding the highest levels of stigma and the least favorable attitudes toward seeking help, followed by women and then nonbinary individuals. In Study 1, this pattern emerged robustly across measures of Social Distance, Stereotypes, and Attitudes Toward Seeking Help. The gendered differences were consistently reproduced across models, iterations, and measures under the present prompting setup.

In Study 2, the comparison with the HMS data confirmed that the AI estimations captured the relative gender differences observed in human responses for Personal Stigma and Self Stigma. In both the human data and the AI estimations, men had the highest levels of these stigma types. This alignment indicates that the AI models' estimates tracked the direction of real-world group differences for these constructs.

However, the analysis also discovered substantial discrepancies in the absolute magnitude of stigma between AI estimations and human responses. The AI models underestimated Personal Stigma (stigmatizing attitudes toward others) and Perceived Stigma (perception of others' attitudes) while significantly overestimating Self Stigma (internalized shame). These systematic inaccuracies and biases indicate that while the AI models recognize that some gender groups are more likely to hold certain attitudes than others, they misrepresent and distort how strongly those attitudes are held. These discrepancies also suggest that the AI models underestimate social forms of stigma while overestimating internal stigma.

Underestimating social forms of stigma while overestimating internal stigma may be a direct consequence of the LLMs' training architecture. Although the architectures and development processes of LLMs have become increasingly opaque amid the growing dominance of proprietary AI models including GPT models used in this research (Liao & Vaughan, 2024), here we would like to point out factors that may plausibly account for these patterns. On one hand, the model's "safety alignment," likely achieved through Reinforcement Learning from Human Feedback (RLHF), may penalize the generation of content that appears judgmental or attributes strong negative biases to certain social groups (Ganguli et al., 2023; Ouyang et al., 2022). When prompted to estimate Personal or Perceived Stigma, this alignment may act as an internal guardrail, forcing the model toward more socially desirable, attenuated responses and thus underestimating the true prevalence of these social stigmas. This possibility is especially relevant to Personal Stigma, where AI estimates were concentrated near the lower end of the response scale. This floor-adjacent pattern may partly reflect the model's tendency not to attribute stigmatizing attitudes to certain social groups, rather than a calibrated estimation of low stigma. On the other hand, the model's foundational pre-training corpus is vast and likely abundant with first-person narratives of mental distress, such as those from social media platforms like Twitter and Reddit (Longpre et al., 2024), which often emphasize personal feelings of shame, guilt, and inadequacy. When prompted for Self Stigma, the model may draw heavily on these pervasive narratives, which are less directly moderated by the socially-focused internal guardrails. This might create an internal tension in AI models, where they simultaneously present society as more accepting than it is, and the individual as more internally stigmatized than they are.

The findings have important implications for the development and deployment of AI applications in mental health, such as conversational agents and diagnostic tools. The presence of consistent, gendered representations in the GPT-based AI models highlights the risk that these technologies may perpetuate and amplify existing stereotypes (Carr, 2020; Timmons et al., 2023). For instance, if an AI-powered chatbot operates under the internalized assumption that men are resistant to seeking help, it may not prioritize suggesting help-seeking behaviors to male users or may inform them that men typically do not seek professional help. Also, the misalignments between AI estimations and human perceptions highlight the risk of AI acting as a distorted social portrayal. As users increasingly turn to AI for information and support, these models can shape their understanding of how mental illness is viewed in society. The systematic underestimation of Personal and Perceived Stigma might lead users to believe that societal attitudes are more accepting than they truly are, potentially leaving them unprepared for real-world discrimination or invalidating their experiences (Shen et al., 2024). The exaggeration of Self Stigma, on the other hand, could intensify users' feelings of isolation and shame, creating additional barriers to help-seeking (Carr, 2020). For example, when users seek advice, interactive AI agents influenced by an exaggerated depiction of Self Stigma might frame guilt, shame, or personal weakness as common responses to mental illness. This could lead the AI to deprioritize help-seeking suggestions or cause users to perceive such stigmatization as normal and unavoidable.

These risks are closely tied to questions of AI literacy. If AI systems can shape users' perceptions of what others believe about mental illness, then it becomes critical that users and practitioners understand how to interpret such outputs. Users may interpret AI-generated statements about how “people” or specific groups think about mental illness as descriptive facts about society, when in fact such outputs are model-generated representations that may only partially align with real-world data. Improving AI literacy in mental health contexts therefore requires teaching users and practitioners to treat these outputs as fallible approximations that should be interpreted cautiously, especially when they concern stigma, help-seeking, or socially sensitive group differences.

The findings also emphasize the need for rigorous, context-specific validation of AI models against real-world data. As the present research demonstrates, even advanced models can exhibit clear inaccuracies and biases that are not immediately apparent without direct comparison to human responses. Ensuring the fairness and equity of AI in mental health requires moving beyond general benchmarks to evaluating model performance across diverse demographic groups and specific psychological constructs.

This research is not without limitations. First, the analysis focused on an LLM family, OpenAI’s GPT. While these models represent the current state-of-the-art, the observed patterns may not generalize to other models with different architectures or training data. Future research should examine a broader range of LLMs to determine the universality of these findings. Second, the comparison in Study 2 relied on the HMS data based on responses from U.S. postsecondary students. While this population is undergoing increasing mental health difficulties (Mojtabai & Olfson, 2020; Patel & Lewis, 2023), this population is not representative of the general population or other cultural contexts. Related to this limitation, Study 1 and Study 2 used different reference populations in the AI prompts. Therefore, comparisons across the studies should be interpreted as convergence across related but not population-identical contexts. Some differences across studies, including the less consistent Perceived Stigma finding, may partly reflect this difference in reference populations. Also, there may be a temporal mismatch between the data informing the AI models' representations and the 2022-2023 HMS benchmark. Because stigma-related attitudes can shift over time, this temporal gap may have contributed to some of the observed misalignment between AI estimations and human-reported perceptions. Third, the prompting strategy used here was designed for standardized measurement and direct comparison with human survey data. It does not approximate the multi-turn, personalized, emotionally embedded way in which many users interact with conversational AI systems in practice. Accordingly, the present findings should not be taken as a direct estimate of how stigma-related bias would unfold in extended real-world conversations. Lastly, the present research focused specifically on gender and did not examine how multiple aspects of social identity and demographic background, such as race, ethnicity, age, immigration status, or socioeconomic position, may jointly shape AI estimations of mental health perceptions. This narrower focus was intentional. Before addressing more complex intersectional configurations, research must first establish rigorous building blocks by examining individual social dimensions in relation to their theoretical foundations, real-world patterns, and AI-specific implications. We therefore view the present research as an important first step and encourage future work to extend this approach to other individual social dimensions as well. Such building blocks will ultimately support intersectional research that treats each dimension as more than an added variable and more meaningfully examines both the distinct contribution of each dimension and their interactions.

## Conclusion

The current research provides a detailed examination of how advanced AI models based on GPT portray mental health stigma across gender groups. The findings reveal that the models display consistent, gendered representations that reflect traditional stereotypes, accurately capturing the relative patterns of stigma observed in human populations for some constructs but not others. The AI models also exhibit significant discrepancies in the magnitude of stigma, compared to human perceptions in the real world. These results highlight the complex ways in which AI models represent human perceptions and underscore the need for ongoing evaluation, bias mitigation, and transparent validation to ensure the safe and equitable deployment of AI in the sensitive domain of mental health.

**Data Availability**

Replication data, including item-level response distributions and detailed statistical results for both GPT-4 and GPT-5 analyses, are available in the Supplementary Online Material (https://osf.io/gby8m).

Figure 1. AI-Estimated Mental Health Attitudes Across Gender Groups (Study 1)

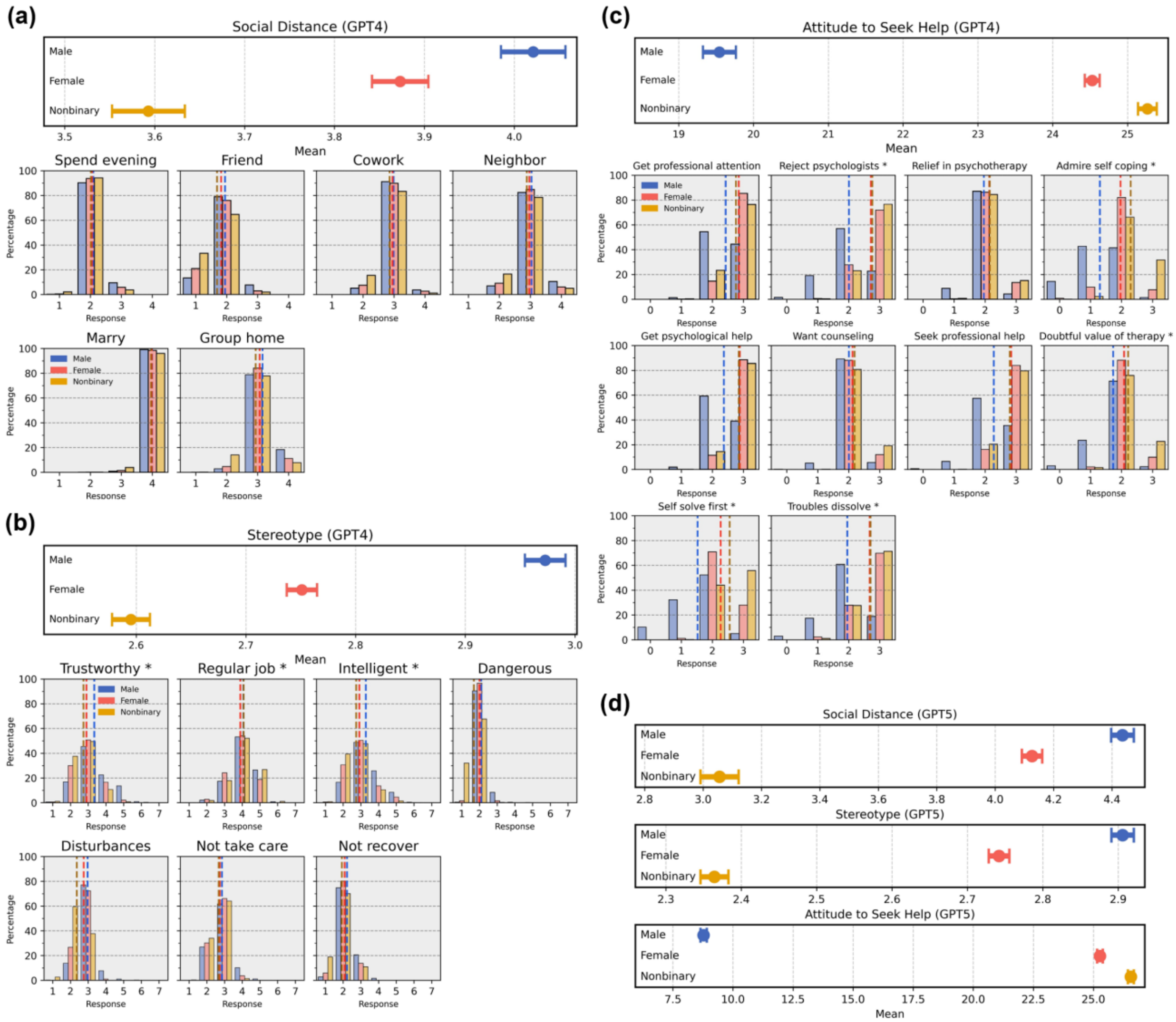

*Note*. The top-level error bar plots of Figures 1a-1c present the overall scores, with the dots indicating the means and the error bars representing the 95% confidence intervals. The bar plots below show the response distribution for each individual item, with the vertical lines indicating the mean response for each gender group. Items with titles marked by an asterisk indicate reverse-coded results and show the corresponding adjusted values. Figure 1d presents the overall scores from GPT-5.

Figure 2. Comparison Between AI-Estimated and Human-Reported Mental Health Stigma (Study 2)

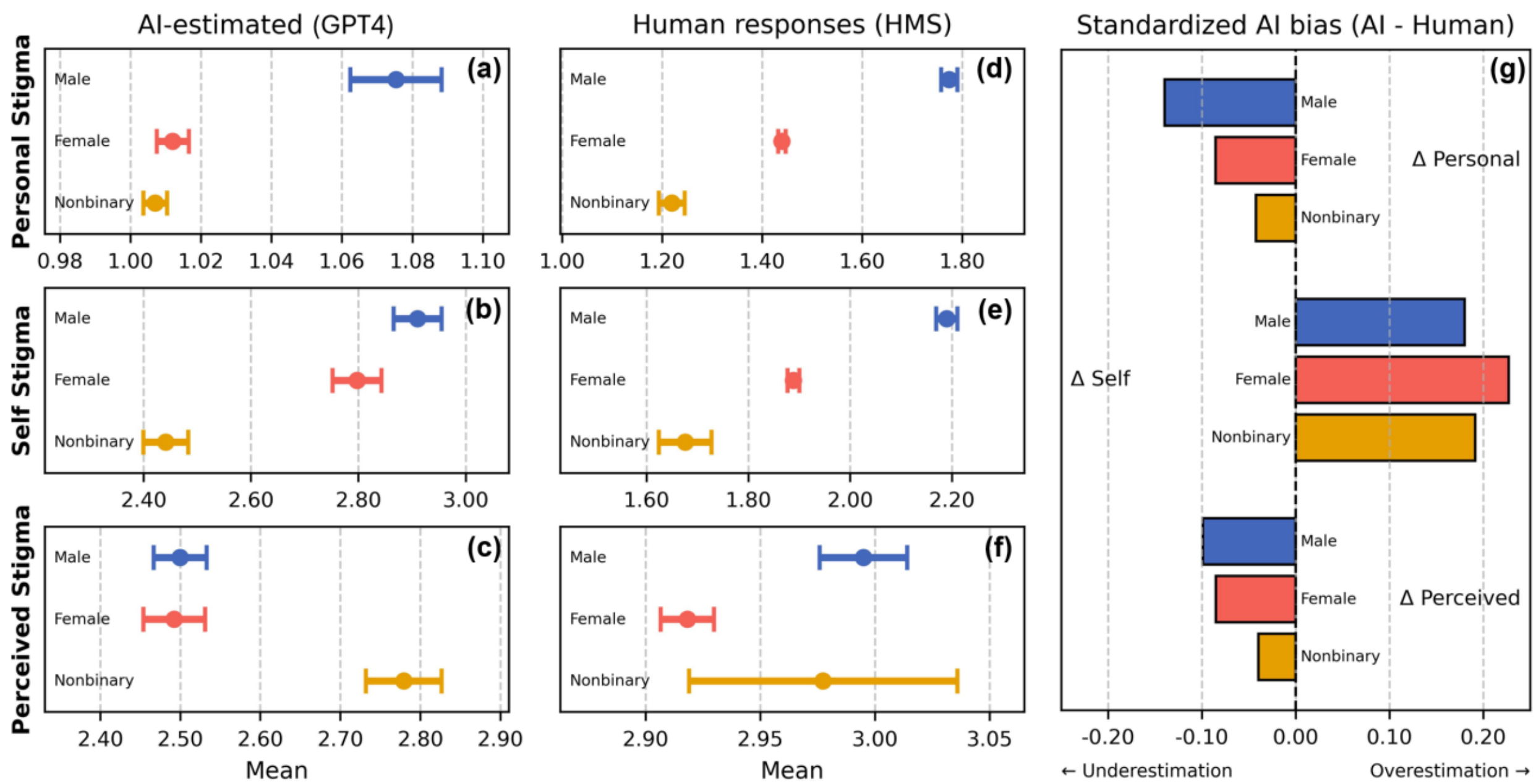

*Note*. The error bar plots (Figures 2a-2f) present the overall scores for each gender group, with the dots indicating the means and the error bars representing the 95% confidence intervals. The bar plot (Figure 2g) presents the range-standardized magnitude and direction of discrepancies between AI estimations and human responses for each measure and gender group.